 \documentclass[final,5p,times,twocolumn]{elsarticle}

\usepackage{amssymb}
\usepackage{graphicx}
\usepackage{lipsum}
\usepackage{subfig}
\usepackage{xcolor}
\usepackage{soul}

\journal{Physics Letters B}

\begin{document}

\begin{frontmatter}



\title{Constraining capture cross sections using proton inelastic scattering as a surrogate reaction}
 \author{Aaina Thapa\fnref{label1}}
\ead{aaina1@llnl.gov}
\author{Jutta Escher\fnref{label1}}
\author{Emanuel Chimanski \fnref{label2}}
\author{Oliver Gorton\fnref{label1}}
\author{Marc Dupuis \fnref{label3,label4}}
\author{Eun Jin In\fnref{label5}}
\author{Shuya Ota \fnref{label2}}
\author{Sophie P\'{e}ru \fnref{label3,label4}}
\author{Walid Younes\fnref{label1}}
\cortext[cor1]{}
\affiliation[label1]{organization={Lawrence Livermore National Laboratory (LLNL)},
            city={Livermore},
            postcode={94550}, 
            state={California},
            country={USA}}
            
\affiliation[label2]{organization={Brookhaven National Laboratory},
            city={Upton},
            postcode={11973}, 
            state={New York},
            country={USA}}
            
\affiliation[label3]{organization={CEA},
            city={DAM, DIF},
            state={F-91297 Arpajon},
            country={France}}
            
\affiliation[label4]{organization={Universit\'{e} Paris-Saclay, CEA},
            city={Laboratoire Mati\'{e}re sous Conditions Extr\^{e}mes},
            state={91680 Bruy\'{e}res-Le-Ch\^{a}tel},
            country={France}}
            
\affiliation[label5]{organization={Louisiana State University},
            city={Baton Rouge},
            postcode={70802}, 
            state={Louisiana},
            country={USA}}

\begin{abstract}
The surrogate reaction method is an alternative to direct measurements of compound nuclear reaction cross sections. We introduce theory tools for extracting capture cross sections from experiments that use proton inelastic scattering as a surrogate reaction mechanism. This makes it possible to constrain compound nucleus decay models which are typically the largest source of uncertainty in capture cross section calculations. This letter describes the theory developments that were used to simultaneously infer $^{89}$Y$(p,\gamma)$ and $^{89}$Zr$(n,\gamma)$ cross sections from $^{90}$Zr$(p,p'\gamma)$ surrogate measurements.

\end{abstract}






\end{frontmatter}




\section{Introduction}
\label{introduction}
Knowledge of capture reactions is crucial for nuclear technology applications and for understanding nucleosynthesis in stellar environments \cite{Capture-1, Capture-2, Capture-3, mumpower-2016}. Accurate neutron capture cross sections are needed across a wide range of isotopes, yet direct measurements face severe limitations for reactions involving short-lived or highly radioactive nuclei, as neither can be readily produced as targets. Radioactive beam facilities have substantially extended the experimental reach to nuclei away from stability, however, direct measurements of the neutron-induced reactions on these isotopes are still out of reach. On the other hand, theoretical predictions using the Hauser-Feshbach framework suffer from significant uncertainties in the relevant decay models~\cite{RIPL,HF-problems,NLD-GSF-UQ-1,NLD-GSF-UQ-3, TENDL, mumpower-2016}. Given these challenges, it is important to develop and test indirect methods for determining capture cross sections. In recent years, it has been demonstrated that $(p,d)$ and $(d,p)$ experiments can successfully be used as surrogate reactions to constrain neutron-induced reactions of interest \cite{escher-review2025, Andrew-2019, escher-2018}. More recently, $^{208}$Pb$(p,p'\gamma)$ measurements, performed at the experimental storage ring (ESR) of the GSI/FAIR facility, were used to extract the neutron capture cross sections for  $^{207}$Pb~\cite{Pb208-ref}. Inelastic scattering is also a promising surrogate mechanism for constraining $(n,2n)$ reactions given that a wealth of giant resonance studies have shown that compound nuclei can be produced that way at high excitation energies. The goal of this work is to introduce several new theory tools to enable the extraction of capture cross sections from proton inelastic scattering surrogate experiments. These extensions include the calculation of two-step processes in populating the target nucleus in the proton inelastic scattering surrogate mechanism, Markov Chain Monte Carlo (MCMC) parameter inference for the level density and gamma-ray strength function parameters, and accounting for the impact of partial-width fluctuations on the $\gamma$-emission probability due to the low-density of states available in the particle emission channels. We apply these tools to existing $^{90}$Zr$(p,p'\gamma)$ data to constrain the neutron capture cross section for $^{89}$Zr isotope which has a half-life of~$\approx 3.72$ days. We simultaneously determine the known proton capture cross section for $^{89}$Y, to test our approach. Improvements in this mass region are essential for constraining the neutron capture cross sections for isotopes located at neutron shell closures, $s$-process branching points, and future applications to exotic isotopes~\cite{annual-review-2023,sprocess-1,sprocess-2,Lugaro_2003}.
\section{Surrogate Reaction Theory}
Theoretical prediction for compound neutron-capture cross sections, based on the Hauser-Feshbach theory~\cite{HF-original}, given by 
\begin{eqnarray}
    \lefteqn{\sigma_{(n,\gamma)}(E_n)} \nonumber \\
    &=& \sum_{J,\pi} \sigma_{\rm{CN}}(E_{\rm{ex}},J, \pi) G_{\gamma} (E_{\rm{ex}}, J, \pi) W_{n,\gamma}(E_{\rm{ex}}, J, \pi), 
     \label{Eq2}
\end{eqnarray}
often suffer from orders-of-magnitude uncertainties due to poorly constrained level-density and gamma-ray strength functions that constitute the $\gamma$-decay probability, $G_{\gamma} (E,J,\pi)$~\cite{TENDL,mumpower-2016,NLD-GSF-UQ-1,NLD-GSF-UQ-3}. $\sigma_{\rm{CN}}$, the  cross section for forming the compound nucleus by capture can be calculated with reasonable accuracy by one of the available nucleon-nucleus optical potentials~\cite{koning2003,bauge2001,Pruitt-2023}. $W_{n,\gamma} (E_{\rm{ex}}, J, \pi)$ are width-fluctuation corrections which are well-studied~\cite{WF-2003, WFC-original1, WFC-original0}. Here, $E_{\rm{ex}}, J, \pi$ are the excitation energy, total angular momentum and parity of the compound nucleus, respectively, $E_n =\frac{(A_{\rm{CN}} + 1)}{A_{\rm{CN}}} (E_{\rm{ex}}-S_n)$ with $A_{\rm{CN}} = 90$ being the mass number, and $S_n = 11.97$ MeV the neutron separation energy of $^{90}$Zr.  In this work, we determine the $^{89}$Zr$(n,\gamma)$ and $^{89}$Y$(p,\gamma)$ cross sections, as both involve the decay of $^{90}$Zr. For the latter, expressions analogous to Eqn.~(\ref{Eq2}) apply, with the proton separation energy, $S_p=$ 8.35 MeV. \\

In surrogate reaction applications, the key idea is to constrain the models describing the decay of the compound nucleus - in our case $^{90}$Zr. This is accomplished by exciting the nucleus via direct proton inelastic scattering in a surrogate experiment and observing the $\gamma$-decay through discrete transitions from states $i$ to $f$. Proton-$\gamma$ coincidence probabilities, $P_{p'\gamma}(i \rightarrow f)$, are measured, which can be expressed as, 
\begin{equation}
    P_{p'\gamma}^{i\rightarrow f} (E_{\rm{ex}})= \sum_{J, \pi} F_{pp'}^{\rm{CN}} (E_{\rm{ex}}, J, \pi) G_{\gamma} (E_{\rm{ex}}, J, \pi)
    \label{Eq1}
\end{equation}
The surrogate reaction data used in this work are from $^{90}$Zr$(p,p' \gamma)$ coincidence probabilities measurements,  previously published~\cite{Ota-2015}, using the proton beam (with beam energy 28.56 MeV) at the K150 Cyclotron facility at Texas A\&M University. No new analysis of the published experimental data was done for this work. \\

The competition between $\gamma$-emission and other decay channels open at excitation energies $E_{\rm{ex}} \approx 11$- $13$ MeV leads to a characteristic fall off of the coincidence probabilities around the neutron separation energy. This feature can be used to constrain the $\gamma$-decay probability, $G_{\gamma} (E,J,\pi)$ which in turn restricts the compound capture cross sections using Eq.~(\ref{Eq2}). To constrain the level density and gamma-ray strength functions needed to predict the desired capture cross sections we have to (i) calculate the spin-parity population $F_{p,p'}^{\rm{CN}}$ in $^{90}$Zr following the proton-inelastic scattering reaction, and (ii) model the decay of the populated $^{90}$Zr nucleus. The decay of $^{90}$Zr depends on the level density and gamma-ray strength function in $^{90}$Zr, and the low-lying states in the neighboring $^{89}$Zr and $^{89}$Y nuclei. \\

\subsection{Calculating the spin-parity population $F_{p,p'}^{\rm{CN}} (E_{\rm{ex}}, J, \pi)$}
\begin{figure}
\centering 
\includegraphics[width=0.5\textwidth]{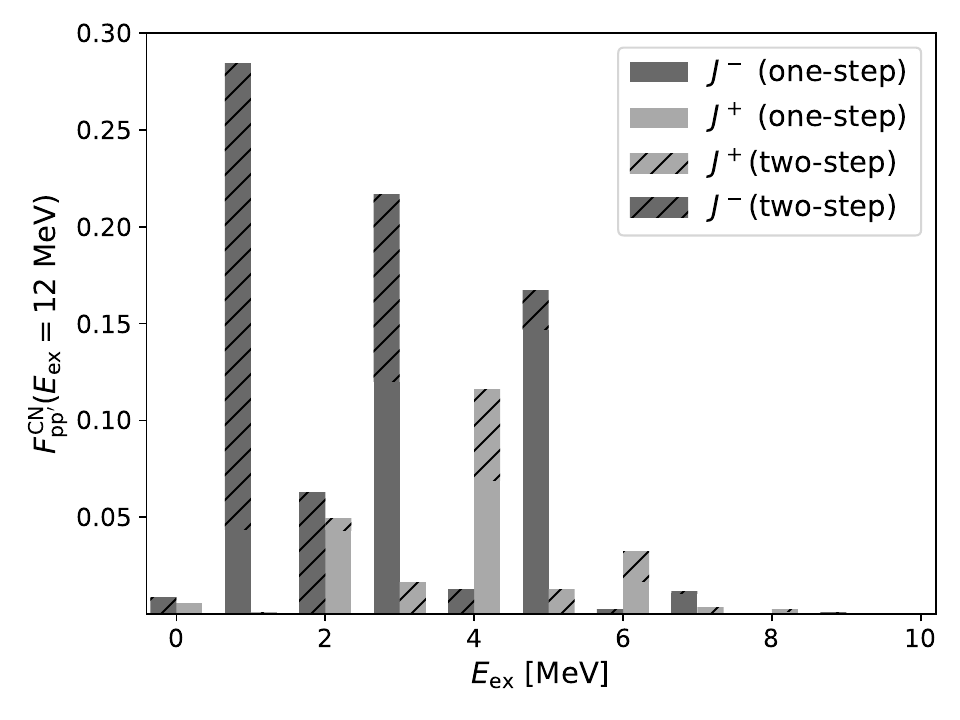}	
\caption{The spin-parity population, $F_{p,p'}^{\rm{CN}} (E_{\rm{ex}}, J, \pi)$, from both one-step inelastic scattering (solid bars) and two-step $(p,d)(d,p')$ process (hatched bars). It shows that the 1- states are most likely to be populated in the $^{90}$Zr nucleus near $S_n$, followed by the $3-$ and $5-$ states. The angular momentum quantum number for the excited states is shown on $x$-axis and the $y$-axis gives the probability of populating a particular $J,\pi$ state at 12 MeV excitation energy in $^{90}$Zr. } 
\label{Spinpop}%
\end{figure}

The nuclear decay via $\gamma$-emission is sensitive to the starting spin and parities of the compound nucleus. The spins and parities populated by a surrogate reaction are typically different from those occurring in compound nucleus formation by nucleon capture~\cite{SPP-2020,SPP-2016,escher-2012-review,Boutoux-2012,escher-2010-1,escher-2010,Chiba-2010, Forssen-2007}. Thus, accurate tuning of the level density and the gamma-ray strength function models to reproduce the coincidence probabilities relies on the theoretical description of $F_{pp'}^{\rm{CN}}$, also known as the spin-parity population, which is a function of the excitation energy of the target nucleus, $E_{\rm{ex}}$, 
\begin{eqnarray}
\lefteqn{F_{p,p'}^{\rm{CN}}(E_{\rm{ex}}, J, \pi)}  \\ \nonumber
&= &\frac{\sum_{E_{J^\pi}}L(E_{\rm{ex}}, E_{J^\pi})\int_{\theta_{\rm{min}}}^{\theta_{\rm{max}}} \sigma_{pp'}( E_{J^\pi}, \theta) {\rm{sin}} \theta d\theta}{\sum_{J',\pi'}  \sum_{E_{J'^{\pi'}}}L(E_{\rm{ex}}, E_{J'^{\pi'}}) \int_{\theta_{\rm{min}}}^{\theta_{\rm{max}}}\sigma_{pp'}( E_{J'^{\pi'}}, \theta) {\rm{sin}} \theta d\theta}.
\end{eqnarray}
\begin{figure*}[h!]
\centering
    \subfloat{\includegraphics[width=0.25\textwidth]{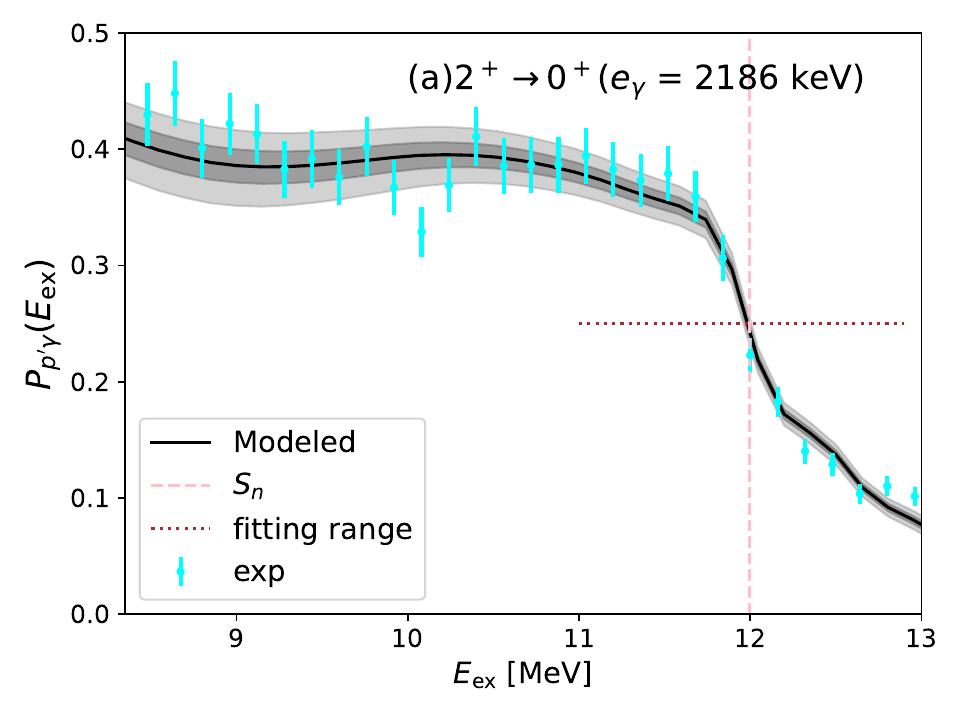}} 
    \subfloat{\includegraphics[width=0.25\textwidth]{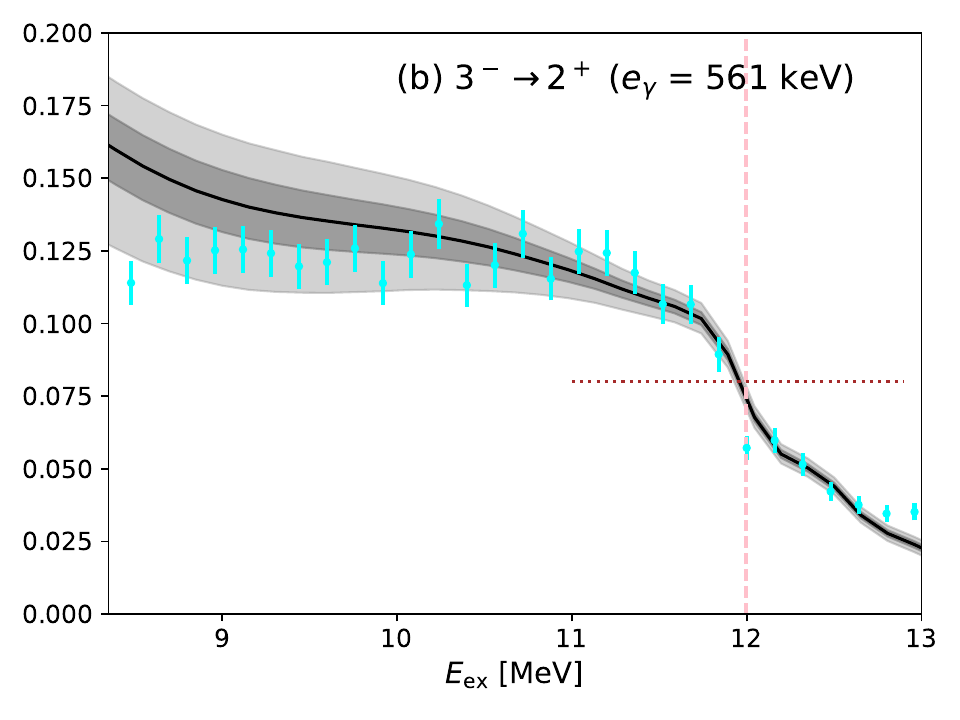}} 
    \subfloat{\includegraphics[width=0.25\textwidth]{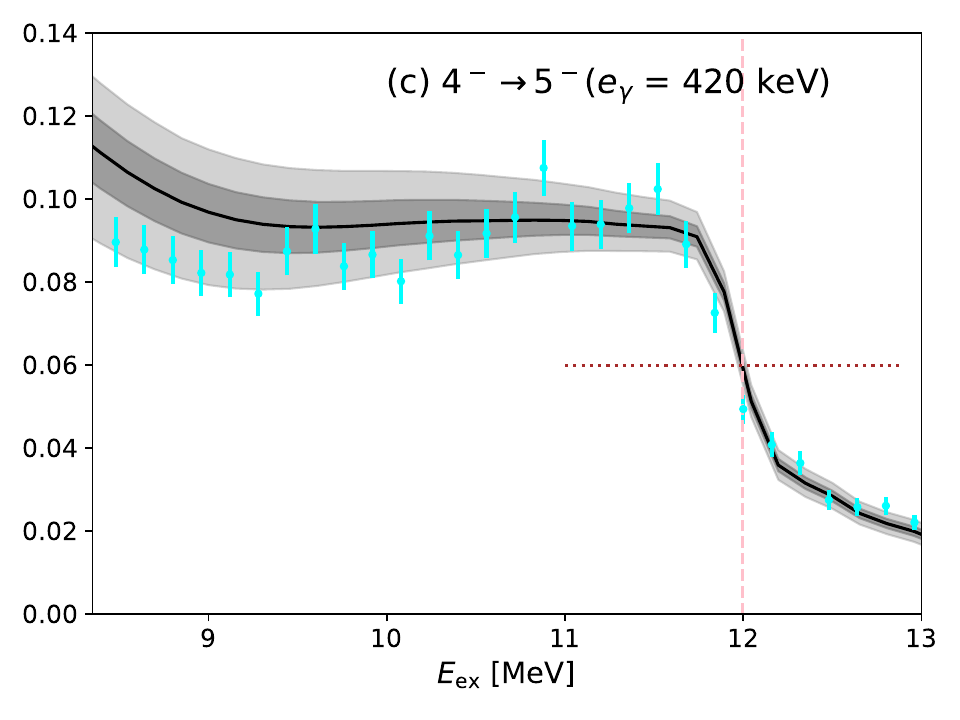}}
    \subfloat{\includegraphics[width=0.25\textwidth]{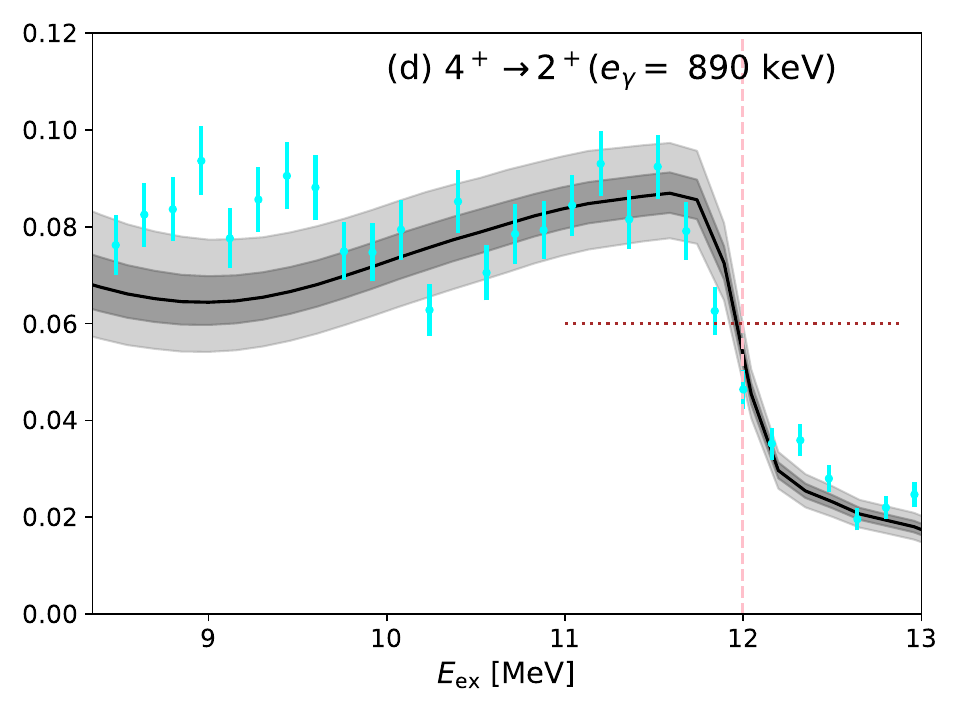}}
    \caption{Comparison between modeled and measured coincidence probabilities, $P_{p'\gamma} (E_{\rm{ex}})$, as a function of the excitation energy, $E_{\rm{ex}}$, for discrete transition in $^{90}$Zr (indicated on each subplot). The black curve is the median, dark grey and light grey are the $68\%$ and 98\% confidence bands of the MCMC posterior, respectively. The vertical dashed-line marks the neutron separation energy $S_n$ for $^{90}$Zr and the horizontal dotted line shows the excitation energy range for which fitting is performed in the MCMC run. }
\label{Ppgs} 
\end{figure*}
Here, $\sigma_{pp'}( E_{J^\pi}, \theta)$ are the angular differential cross sections for discrete excited states at energies $E_{J^\pi}$  with spin $J$ and parity $\pi$ in $^{90}$Zr, calculated from direct one-step inelastic scattering and a two-step process which involves the intermediate pick-up and drop-off of a neutron by the incoming proton. The differential cross sections are integrated from 25 to 60 degrees scattering angle in the laboratory frame to match the angular range of the measured coincidence probabilities. The impact on inelastic scattering due to fragmentation of the discrete excited states from particle-hole excitations beyond what is already included in calculating the one- and two-step excitation spectra is accounted for using Lorentzian spreading, $L (x,x_0) = \frac{1}{\pi} \frac{\Gamma/2}{(x-x_0)^2 + (\Gamma/2)^2} $,  of the angle-integrated cross sections using the energy-dependent width, $\Gamma$~\cite{Smith-Wambach1988, damping-2,damping-1}. \\

To generate the population of the $^{90}$Zr excited states due to one-step direct inelastic scattering from its ground state, states of $^{90}$Zr were calculated using the D1M-Gogny interaction with Hartree-Fock-Bogoliubov (HFB) plus Quasi-Random-Phase Approximation (QRPA) many-body methods~\cite{Chimanski-2025,Chimanski-2022,Peru2014,Peru-2011}. The QRPA transition densities folded with the JLM-B nucleon-nucleon interaction, were used to produce the coupling potentials~\cite{jlm-1977, jlmb-1998, bauge2001}. These microscopic coupling potentials were used as input in the coupled-channels code FRESCOX~\cite{Thompson1988} to perform differential inelastic scattering cross section calculations for all natural parity QRPA states up to 20 MeV excitation energy and maximum total angular momentum $J_{\rm{max}}=$ 10 under the distorted Born wave approximation (DWBA). This approach has been shown to do well for inelastic scattering differential cross sections of low-lying excited states~\cite{dupuis2019, dupuis2017, 2018-inelastic-jlm}. We presented the agreement between our implementation and the experimental inelastic scattering data for the $^{90}$Zr nucleus in \cite{thapa2023}. \\

For the present surrogate study, we are interested in the population of excited states close to the neutron separation energy which can also be the outcome of a two-step process that involve intermediate pick-up and drop-off of a neutron in $^{90}$Zr. The detector counting outgoing protons in the surrogate measurements, corresponding to these high excitation energies of $^{90}$Zr, does not distinguish between one- and two-step contributions. The importance of accounting for two-step processes in the spin-parity populations was first highlighted in \cite{escher-2018} for $(p,d)$ surrogate reactions. So, in addition to the microscopically calculated one-step populations, we account for the two-step $(p,d)(d,p')$ contributions to the populated spin and parities. This contribution is calculated using two-step DWBA combining (a) a neutron pick-up which leaves behind a $^{89}$Zr nucleus, described as a hole in the occupied single particle orbitals of $^{90}$Zr, with (b) neutron drop-off to an empty single particle orbital to form an excited $^{90}$Zr nucleus with an outgoing proton. The single particle orbital energies were taken from a combination of available experimental data~\cite{hole_states_data1, hole_states_data2} and dispersive optical potential work~\cite{bespalova-2015}. \\

\begin{figure*}[h!]
\centering 
\subfloat{\includegraphics[width=0.4\textwidth]{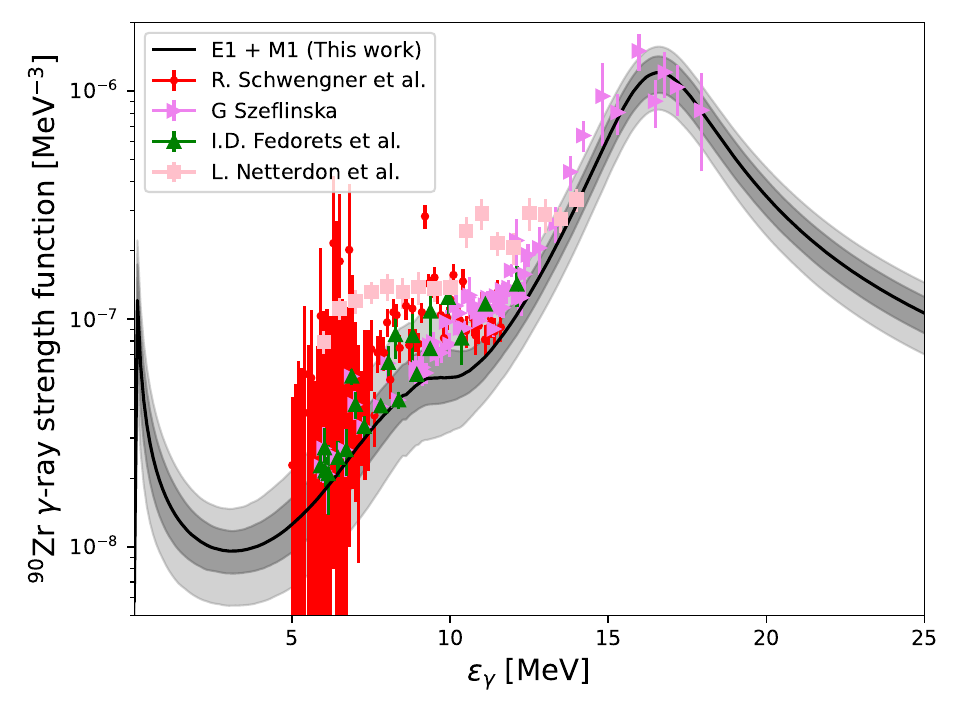}}	
\subfloat{\includegraphics[width=0.4\textwidth]{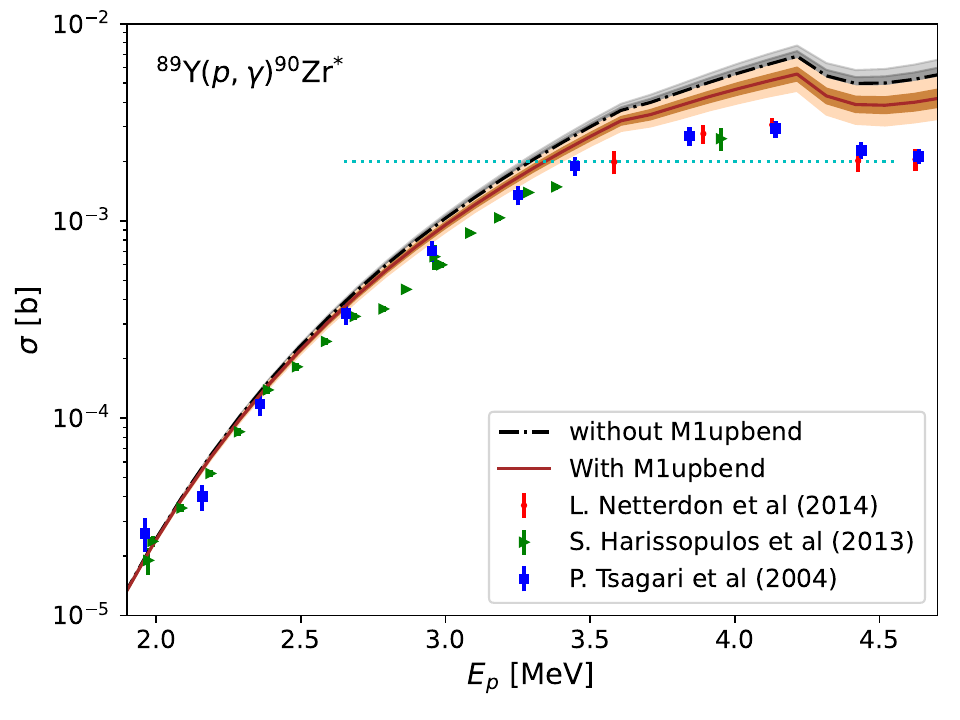}}
\caption{(a) The $E$1 $+ M$1 GSFs obtained from the MCMC fit to the surrogate data (black curve with 68\% and 98\% grey uncertainty bands) compared to data from references \cite{netterdon-2015, Fedorets_2013, PDR-2008, GS-1979}. (b) Proton capture cross sections obtained from the surrogate fit with 68\% and 98\% light brown and peach bands, respectively. The brown (black) solid curve and bands were obtained with the models described in section~\ref{fit}. The black curve (along with the dark grey 68\% and light grey 98\% uncertainty bands) is the result of a fit without including an M1 upbend in fitting NLD and GSF to the surrogate data. Dots show experimental data from references~\cite{netterdon-2015, Y89pg-2, Y89pg-3}.} 
\label{Y89pg}
\end{figure*}
Around the neutron separation energy, the resulting spin-parity population $F_{pp'}^{\rm{CN}}(E_{\rm{ex}}, J, \pi)$, calculated for $E_{\rm{ex}} = 0.1$~to~20 MeV, was found to vary slowly with excitation energy. The result for $E_{\rm{ex}} = 12$ MeV is shown in Fig.~\ref{Spinpop}. The dark grey bars show negative-parity contributions and light grey bars mark the contributions from positive-parity excited states. The contributions coming from one-step and two-step processes to the spin-parity population are indicated by solid and hatched bars, respectively. It is clearly seen that the addition of two-step processes have a significant impact on the spin-population around neutron separation energy. The population of unnatural parity states in Figure~\ref{Spinpop} is purely an outcome of including two-step contributions. Since the JLM-B single-folding model does not contain parity-violating nucleon-nucleon interaction terms, the model cannot couple the $0^+$ ground state to unnatural parity excited states.   \\

\subsection{Constraining the nuclear level density (NLD) and the $\gamma$-ray strength functions (GSF)}
\label{fit}
The Hauser-Feshbach decay calculations, for which a modified version of the reaction code YAHFC~\cite{YAHFC} was used, start with the calculated energy-dependent spin-parity populations of $^{90}$Zr,  $F_{pp'}^{\rm{CN}}(E_{\rm{ex}}, J, \pi)$. The proton separation energy $S_p = 8.350$ MeV is below the neutron separation energy, $S_n = 11.97$, for $^{90}$Zr. Due to the Coulomb barrier, the proton channel opens at around 10.2 MeV excitation energy. In the energy range where the $P_{p'\gamma}(E_{\rm{ex}})$ fall-off, $11  \lesssim E_{\rm{ex}} \lesssim 13$, at most two excited states in the $^{89}$Zr daughter nucleus can be reached by neutron emission, and up to eight excited states in $^{89}$Y are accessible by proton emission. As a result, both the proton- and neutron-emission partial widths deviate from the mean of the Porter-Thomas distribution~\cite{PT-original}, as is known from $\beta$-delayed neutron emission, see~\cite{gorton-2025, PT-dicebox, PT-exact} and references therein. Correcting for this effectively increases the $\gamma$-decay probability of the compound nucleus relative to the Hauser-Feshbach estimate, with the latter being applicable when the particle emission channels have a large number of final states available. The model calculations in the MCMC fits include this effect. The detailed analysis of the impact of these corrections on surrogate reaction analysis will be presented in a separate paper~\cite{PT-WIP}.  \\

$E$1 and $M$1 transitions, the lowest-order transitions in the multipole expansion of electromagnetic radiation, are known to dominate the $\gamma$-decay. We use the Kopecky-Uhl $E$1 strength function~\cite{Kopecky-1990}, which includes a Lorentzian giant-dipole resonance (GDR) resonance with a temperature-dependent width plus a temperature-dependent strength for small $\gamma$-ray energies, and add an enhancement to represent a Pygmy resonance. For the $M$1 de-excitation strength function a standard Lorentzian with a low-energy upbend is included. We choose the composite level density model by Gilbert and Cameron which combines the constant temperature model at low-excitation energies with the back-shifted Fermi-gas model at high excitation energies \cite{gilbert-1965}. \\

To constrain the decay probabilities, $G_{\gamma} (E_{\rm{ex}}, J, \pi)$, of the compound nucleus, we fit 14 parameters in the NLD and GSF models discussed above to reproduce the experimental coincidence probabilities. For parameter fitting we use the LLNL code COMMCAS~\cite{commcas-2025, gorton2020neutron}, a parameter inference tool which uses affine-invariant MCMC sampling~\cite{goodman2010ensemble}.  The resulting posterior probability distribution of the fitted parameters is statistically consistent (via Bayes’ theorem) with both the experimental coincidence probabilities (including uncertainties) and our prior knowledge of the parameters. The sampling consisted of an ensemble of 200 walkers and 5000 iterations. The prior was chosen after performing a sensitivity study by varying each individual parameter and observing its impact on the coincidence probabilities. Additionally, in choosing the $E$1-GDR parameters, the systematics for this mass region and measured $E$1 photon strength functions were considered. Agreement between the posterior distribution of coincidence probabilities and available data is shown in Figure~\ref{Ppgs}. 

\section{Results}
Before turning to a discussion of the results, we would like to emphasize that in our MCMC approach, all parameters are fit simultaneously, yielding a fully correlated posterior distribution consisting of the NLD and $E$1 and $M$1 strength function parameters. These can subsequently be sampled to calculate proton and neutron capture cross sections, as well as other quantities of interest, such as neutron resonance spacings, radiative widths, and GSFs. \\

In Figure~\ref{Y89pg}~(a), we compare resulting $E$1$+M$1 strength functions plotted as a function of the $\gamma$-ray energy, $\epsilon_\gamma$, to experimental data from references~\cite{netterdon-2015, Fedorets_2013, PDR-2008, GS-1979}. The main peak, at $e_\gamma = 16.9 \pm 0.1$ MeV is the giant dipole resonance, the smaller peak in the $e_\gamma = 7.5$-$10.5$ MeV region is a combination of an $E$1 pygmy dipole resonance (PDR) and a $M$1 spin-flip resonance, and the low-energy upturn has $M$1 character. The PDR centroid is found to be at 9.2$\pm$0.49 MeV and the spin-flip $M1$-resonance is centered at 8.6$\pm$0.57 MeV. The peak energies for both resonances are in agreement with results from other measurements~\cite{PDR-2012, PDR-2008, M1-2013, M1-data-1980}. The magnitude of our extracted GSF is slightly lower but still in agreement with the results from these alternative measurements, except for the Netterdon et al result~\cite{netterdon-2015}, which exhibits more structure and strong enhancements. No definite information is available for the enhancement at very low-energies. We therefore conducted a second fit,  proceeding as outlined above but omitting the low-energy enhancement. We find that including the enhancement improves the reproduction of the shape and magnitude for some coincidence probabilities (comparison not shown).\\

We obtain an average $s$-wave radiative width, $\Gamma_\gamma=195\pm50$ meV, and an $s$-wave neutron resonance spacing, $D_0=38\pm$10 eV from our posterior probability distribution. There are no known experimental values for $\Gamma_\gamma$ and $D_0$. \\

By sampling the posterior of the NLD and GSF parameter distribution, we predict the $^{89}$Y$(p,\gamma)$ cross section. A comparison to experimental results~\cite{netterdon-2015, Y89pg-2, Y89pg-3} is shown in Figure~\ref{Y89pg}~(b). The brown and the black solid curves show the $^{89}$Y$(p,\gamma)$ cross sections sampled from the two different MCMC posteriors obtained with and without a low-energy upbend in the $M$1 strength functions during the MCMC runs, respectively.  We find that the low-energy behavior of the $M$1 strength function affects our predicted $^{89}$Y$(p,\gamma)$ cross section. Proton capture cross sections are known to be also sensitive to the nucleon-nucleus optical potential~\cite{OMP-pg-senstivity} used to calculate $\sigma_{\rm{CN}}$ in Eqn.~(\ref{Eq2}). This sensitivity is not reflected in our uncertainty bands, but can be estimated by sampling the KDUQ optical potential~\cite{Pruitt-2023} while keeping the level density and gamma-ray strength function fixed. We find about $10$\% (5\%) added uncertainty in the proton-capture cross section above (below) 4 MeV proton incident energy. \\

\begin{figure}[h]
\centering 
	\includegraphics[width=0.5\textwidth]{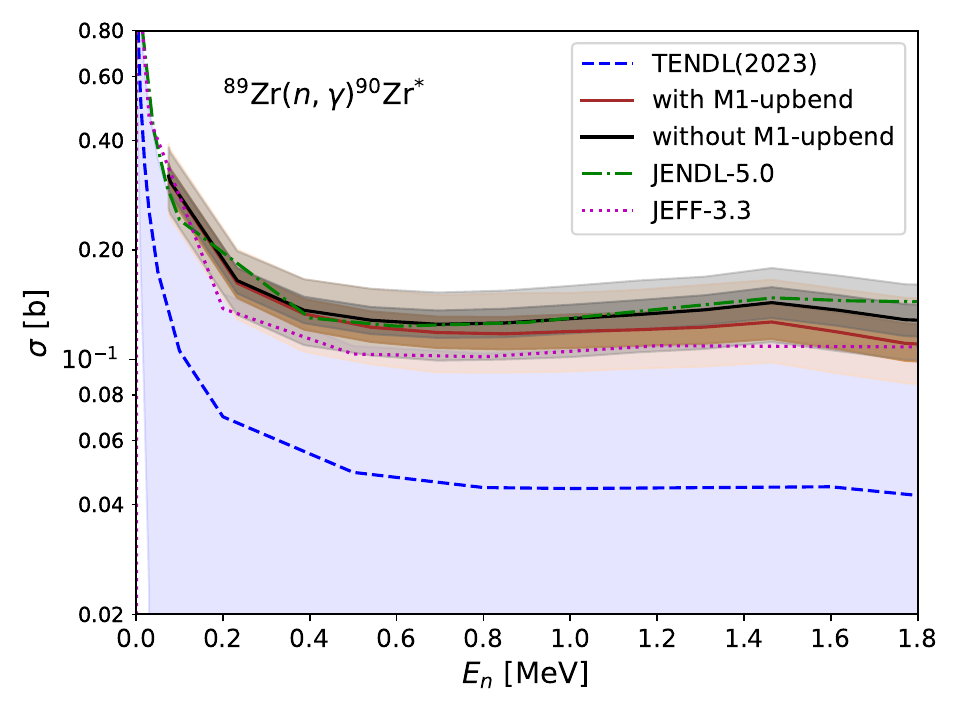}
	\caption{The brown (black) curve shows our extracted $^{89}$Zr$(n,\gamma)$ cross section. The light brown (dark grey) and peach (dim grey) shaded areas give the 68\% and 98\% confidence bands based on the MCMC posterior from fitting to the surrogate data with (without) an upbend in the M1-strength function included. Our results are compared to the nuclear data evaluations JENDL-5.0 (dashed-dotted)~\cite{JENDL5}, JEFF-3.3 (dotted)~\cite{JEFF3} and TENDL-astro-2023 (dashed)~\cite{TENDL-2023}.} 
	\label{Zr89ng}
\end{figure}

Finally, we turn to our result for the unknown neutron capture cross section for the short-lived $^{89}$Zr isotope. Figure~\ref{Zr89ng} shows the constrained neutron capture cross sections (brown solid curve) for $^{89}$Zr in comparison to nuclear data evaluations JENDL-5.0 (dashed-dotted)~\cite{JENDL5}, JEFF-3.3 (dotted)~\cite{JEFF3} and TENDL-astro-2023 (dashed)~\cite{TENDL-2023}. The dark (light) bands show the 68\% (98\%) uncertainty bands based on the probability distribution of parameter values in the MCMC posteriors. Our results are in good agreement with the JEFF-3.3 and JENDL-5.0 evaluations, which rely on regional systematics to constrain the Hauser-Feshbach calculations. The TENDL result is a factor 2.7 lower than our result but has a large uncertainty shown by the blue band. The uncertainty in the TENDL $^{89}$Zr$(n,\gamma)$ cross section arises from combining different recommended formation and decay models in the Hauser Feshbach calculations. The brown (black) solid curves again show the impact of the presence (absence) of an $M1$-upbend in the MCMC fit to the surrogate data. The difference between the predictions from the two models gradually grows as the neutron incident energy increases, however, the medians for both posteriors remain within the 68\% confidence bands of each other.
\section{Summary and Outlook}
We have introduced several new theoretical tools for extracting unkown cross sections from surrogate reaction data. Our developments enable more rigorous uncertainty quantification, application to nuclei in regions of low-level density (near shell-closures and away from stability), and improve the reliability of $(p,p')$ reactions as a surrogate mechanism. We demonstrated our advances by simultaneously extracting proton and neutron capture cross sections from $^{90}$Zr$(p,p' \gamma)$ surrogate data. Determination of spin and parities populated in the proton-inelastic scattering reaction is crucial for constraining the level density and $\gamma$-ray strength functions from the surrogate reaction data. We performed scattering calculations using microscopic nuclear structure inputs. For the first time, two-step contributions, i.e. the intermediate formation and break up of a deuteron, included in calculating the spin-parity populations for the $(p,p')$ surrogate reaction. Another new aspect of this work is the use of MCMC method for level density and gamma-ray strength function parameter optimization. MCMC enables a statistically more rigorous approach to parameter inference and uncertainty propagation, as visualized in both the fits and final capture cross sections. Finally, for the first time, we correct for the impact of Porter-Thomas fluctuations when fitting the NLD and GSF parameters to the surrogate data. This was especially important for $^{90}$Zr $\gamma$-decay because in both the proton- and the neutron-emission channels only very few final states were accessible by particle emission.  \\

The $E$1 and $M$1 strengths for low $\gamma$-ray energies are challenging to measure as they correspond to the high excitation energy region of the nucleus where level-densities are also poorly constrained. A preliminary sensitivity analysis indicates that selected coincidence probabilities might be sufficiently impacted by the low-energy strength of the M1 strength function to constrain
the low-energy behavior of the GSF. In our work, the addition of a low-energy enhancement in the $M$1 strength function improved the agreement with surrogate data and lead to slightly lower $^{89}$Y$(p,\gamma)$ and $^{89}$Zr$(n,\gamma)$ capture cross sections. The predicted cross sections for proton capture by $^{89}$Y is in better agreement with the experimental data when the $M$1 upbend is included in the MCMC fit. The predicted $^{89}$Zr$(n,\gamma)$ cross section, which was found to be in agreement with nuclear data evaluations, exhibits less sensitivity to the low-energy enhancement in the $M$1 strength function. More work is needed to quantify uncertainty in the extracted capture cross sections coming from different low-energy behaviors of the GSF, for example by using the GSFs from the many-body methods like QRPA~\cite{goriely2025} and shell-model~\cite{M1-shell-model-2025,M1-shell-model-2022,M1-shell-model-2018,M1-shell-model-2017,M1-shell-model-2013}.


\section*{Acknowledgements}
A.T. acknowledges Andre Sieverding, Rajesh Ghimire and Richard O. Hughes for insightful discussions. We thank Jeffrey Berryman for helping us assess the uncertainty in the proton-capture cross section due to the parameter variations in the optical model potential. This work was performed under the auspices of the U.S. Department of Energy by Lawrence Livermore National Laboratory under Contract DE-AC52-07NA27344, with partial support from LDRD projects 19-ERD-017 and 20-ERD-030.










\begin{thebibliography}{10}

\bibitem{Capture-1}
B~Alex Brown, Alexandra Gade, S~Ragnar Stroberg, Jutta~E Escher, Kevin Fossez,
  Pablo Giuliani, Calem~R Hoffman, Witold Nazarewicz, Chien-Yeah Seng,
  Agnieszka Sorensen, Nicole Vassh, Daniel Bazin, Kyle~W Brown, Mark~A Caprio,
  Heather Crawford, Pawel Danielewicz, Christian Drischler, Ronald~F
  Garcia~Ruiz, Kyle Godbey, Robert Grzywacz, Linda Hlophe, Jeremy~W Holt, Hiro
  Iwasaki, Dean Lee, Silvia~M Lenzi, Sean Liddick, Rebeka Lubna, Augusto~O
  Macchiavelli, Gabriel Martínez-Pinedo, Anna McCoy, Alexis Mercenne, Kei
  Minamisono, Belen Monteagudo, Petr Navratil, Ryan Ringle, Grigor~H Sargsyan,
  Hendrik Schatz, Mark-Christoph Spieker, Alexander Volya, Remco G~T Zegers,
  Vladimir Zelevinsky, and Xilin Zhang.
\newblock Motivations for early high-profile FRIB experiments.
\newblock {\em Journal of Physics G: Nuclear and Particle Physics},
  52(5):050501, May 2025.

\bibitem{Capture-2}
M.~Arnould and S.~Goriely.
\newblock Astronuclear physics: A tale of the atomic nuclei in the skies.
\newblock {\em Progress in Particle and Nuclear Physics}, 112:103766, 2020.

\bibitem{Capture-3}
Almudena Arcones, Dan~W. Bardayan, Timothy~C. Beers, Lee~A. Bernstein,
  Jeffrey~C. Blackmon, Bronson Messer, B.~Alex Brown, Edward~F. Brown, Carl~R.
  Brune, Art~E. Champagne, Alessandro Chieffi, Aaron~J. Couture, Pawel
  Danielewicz, Roland Diehl, Mounib El-Eid, Jutta~E. Escher, Brian~D. Fields,
  Carla Fröhlich, Falk Herwig, William~Raphael Hix, Christian Iliadis,
  William~G. Lynch, Gail~C. McLaughlin, Bradley~S. Meyer, Anthony Mezzacappa,
  Filomena Nunes, Brian~W. O’Shea, Madappa Prakash, Boris Pritychenko, Sanjay
  Reddy, Ernst Rehm, Grigory Rogachev, Robert~E. Rutledge, Hendrik Schatz,
  Michael~S. Smith, Ingrid~H. Stairs, Andrew~W. Steiner, Tod~E. Strohmayer,
  F.X. Timmes, Dean~M. Townsley, Michael Wiescher, Remco~G.T. Zegers, and
  Michael Zingale.
\newblock White paper on nuclear astrophysics and low energy nuclear physics
  part 1: Nuclear astrophysics.
\newblock {\em Progress in Particle and Nuclear Physics}, 94:1--67, 2017.

\bibitem{mumpower-2016}
M.R. Mumpower, R.~Surman, G.C. McLaughlin, and A.~Aprahamian.
\newblock The impact of individual nuclear properties on $r$-process
  nucleosynthesis.
\newblock {\em Progress in Particle and Nuclear Physics}, 86:86--126, 2016.

\bibitem{RIPL}
R.~Capote, M.~Herman, P.~Obložinský, P.G. Young, S.~Goriely, T.~Belgya, A.V.
  Ignatyuk, A.J. Koning, S.~Hilaire, V.A. Plujko, M.~Avrigeanu, O.~Bersillon,
  M.B. Chadwick, T.~Fukahori, Zhigang Ge, Yinlu Han, S.~Kailas, J.~Kopecky,
  V.M. Maslov, G.~Reffo, M.~Sin, E.Sh. Soukhovitskii, and P.~Talou.
\newblock Ripl – reference input parameter library for calculation of nuclear
  reactions and nuclear data evaluations.
\newblock {\em Nuclear Data Sheets}, 110(12):3107--3214, 2009.
\newblock Special Issue on Nuclear Reaction Data.

\bibitem{HF-problems}
S.~Hilaire and S.~Goriely.
\newblock Towards more predictive nuclear reaction modelling.
\newblock In Jutta Escher, Yoram Alhassid, Lee~A. Bernstein, David Brown, Carla
  Fr{\"o}hlich, Patrick Talou, and Walid Younes, editors, {\em Compound-Nuclear Reactions}, pages 3--15, Cham, 2021. Springer International Publishing.

\bibitem{NLD-GSF-UQ-1}
F.~Pogliano and A.~C. Larsen.
\newblock Impact of level densities and $\gamma$-strength
  functions on $r$-process simulations.
\newblock {\em Phys. Rev. C}, 108:025807, Aug 2023.

\bibitem{NLD-GSF-UQ-3}
M.~Wiedeking and S.~Goriely.
\newblock Photon strength functions and nuclear level densities: invaluable
  input for nucleosynthesis.
\newblock {\em Philosophical Transactions of the Royal Society. A,
  Mathematical, Physical and Engineering Sciences}, 382(2275), 6 2024.

\bibitem{TENDL}
A.J. Koning, D.~Rochman, J.-Ch. Sublet, N.~Dzysiuk, M.~Fleming, and S.~{van der
  Marck}.
\newblock Tendl: Complete nuclear data library for innovative nuclear science
  and technology.
\newblock {\em Nuclear Data Sheets}, 155:1--55, 2019.
\newblock Special Issue on Nuclear Reaction Data.

\bibitem{escher-review2025}
{Escher, Jutta E.}
\newblock The surrogate nuclear reaction method: Concept, recent advances, and
  new opportunities.
\newblock {\em EPJ Web Conf.}, 322:03001, 2025.

\bibitem{Andrew-2019}
A.~Ratkiewicz, J.~A. Cizewski, J.~E. Escher, G.~Potel, J.~T. Harke, R.~J.
  Casperson, M.~McCleskey, R.~A.~E. Austin, S.~Burcher, R.~O. Hughes,
  B.~Manning, S.~D. Pain, W.~A. Peters, S.~Rice, T.~J. Ross, N.~D. Scielzo,
  C.~Shand, and K.~Smith.
\newblock Towards neutron capture on exotic nuclei: Demonstrating
  $(d,p\gamma)$ as a surrogate reaction for
  $(n,\gamma)$.
\newblock {\em Phys. Rev. Lett.}, 122:052502, Feb 2019.

\bibitem{escher-2018}
J.~E. Escher, J.~T. Harke, R.~O. Hughes, N.~D. Scielzo, R.~J. Casperson,
  S.~Ota, H.~I. Park, A.~Saastamoinen, and T.~J. Ross.
\newblock Constraining neutron capture cross sections for unstable nuclei with
  surrogate reaction data and theory.
\newblock {\em Phys. Rev. Lett.}, 121:052501, Jul 2018.

\bibitem{Pb208-ref}
M.~Sguazzin, B.~Jurado, J.~Pibernat, J.~A. Swartz, M.~Grieser, J.~Glorius,
  Yu.~A. Litvinov, J.~Adamczewski-Musch, P.~Alfaurt, P.~Ascher, L.~Audouin,
  C.~Berthelot, B.~Blank, K.~Blaum, B.~Br\"uckner, S.~Dellmann, I.~Dillmann,
  C.~Domingo-Pardo, M.~Dupuis, P.~Erbacher, M.~Flayol, O.~Forstner,
  D.~Freire-Fern\'andez, M.~Gerbaux, J.~Giovinazzo, S.~Gr\'evy, C.~J. Griffin,
  A.~Gumberidze, S.~Heil, A.~Heinz, R.~Hess, D.~Kurtulgil, N.~Kurz,
  G.~Leckenby, S.~Litvinov, B.~Lorentz, V.~M\'eot, J.~Michaud, S.~P\'erard,
  N.~Petridis, U.~Popp, D.~Ramos, R.~Reifarth, M.~Roche, M.~S. Sanjari, R.~S.
  Sidhu, U.~Spillmann, M.~Steck, Th. St\"ohlker, B.~Thomas, L.~Thulliez,
  M.~Versteegen, and B.~W\l{}och.
\newblock First measurement of the neutron-emission probability with a
  surrogate reaction in inverse kinematics at a heavy-ion storage ring.
\newblock {\em Phys. Rev. Lett.}, 134:072501, Feb 2025.

\bibitem{annual-review-2023}
Maria Lugaro, Marco Pignatari, René Reifarth, and Michael Wiescher.
\newblock The s-process and beyond.
\newblock {\em Annual Review of Nuclear and Particle Science}, 73(Volume 73,
  2023):315--340, 2023.

\bibitem{sprocess-1}
S.~Bisterzo, R.~Gallino, F.~Käppeler, M.~Wiescher, G.~Imbriani, O.~Straniero,
  S.~Cristallo, J.~Görres, and R.~J. deBoer.
\newblock The branchings of the main s-process: their sensitivity to $\alpha$-induced reactions on 13C and 22Ne and to the uncertainties of the nuclear network.
\newblock {\em Monthly Notices of the Royal Astronomical Society},
  449(1):506--527, 03 2015.

\bibitem{sprocess-2}
K.~Sonnabend, A.~Mengoni, P.~Mohr, T.~Rauscher, K.~Vogt, and A.~Zilges.
\newblock The $(n,\gamma)$ cross sections of short‐living $s$‐process
  branching points.
\newblock {\em AIP Conference Proceedings}, 704(1):463--472, 04 2004.

\bibitem{Lugaro_2003}
Maria Lugaro, Falk Herwig, John~C. Lattanzio, Roberto Gallino, and Oscar
  Straniero.
\newblock $s$-process nucleosynthesis in asymptotic giant branch stars: A test for stellar evolution.
\newblock {\em The Astrophysical Journal}, 586(2):1305, apr 2003.

\bibitem{HF-original}
Walter Hauser and Herman Feshbach.
\newblock The inelastic scattering of neutrons.
\newblock {\em Phys. Rev.}, 87:366--373, Jul 1952.

\bibitem{koning2003}
A.J. Koning and J.P. Delaroche.
\newblock Local and global nucleon optical models from 1 keV to 200 MeV.
\newblock {\em Nuclear Physics A}, 713(3):231--310, 2003.

\bibitem{bauge2001}
E.~Bauge, J.~P. Delaroche, and M.~Girod.
\newblock Lane-consistent, semi-microscopic nucleon-nucleus optical model.
\newblock {\em Phys. Rev. C}, 63:024607, Jan 2001.

\bibitem{Pruitt-2023}
C.~D. Pruitt, J.~E. Escher, and R.~Rahman.
\newblock Uncertainty-quantified phenomenological optical potentials for
  single-nucleon scattering.
\newblock {\em Phys. Rev. C}, 107:014602, Jan 2023.

\bibitem{WF-2003}
S.~Hilaire, Ch. Lagrange, and A.J. Koning.
\newblock Comparisons between various width fluctuation correction factors for compound nucleus reactions.
\newblock {\em Annals of Physics}, 306(2):209--231, 2003.

\bibitem{WFC-original1}
P.A. Moldauer.
\newblock Statistics and the average cross section.
\newblock {\em Nuclear Physics A}, 344(2):185--195, 1980.

\bibitem{WFC-original0}
P.~A. Moldauer.
\newblock Evaluation of the fluctuation enhancement factor.
\newblock {\em Phys. Rev. C}, 14:764--766, Aug 1976.

\bibitem{Ota-2015}
S.~Ota, J.~T. Harke, R.~J. Casperson, J.~E. Escher, R.~O. Hughes, J.~J.
  Ressler, N.~D. Scielzo, I.~J. Thompson, R.~A.~E. Austin, B.~Abromeit, N.~J.
  Foley, E.~McCleskey, M.~McCleskey, H.~I. Park, A.~Saastamoinen, and T.~J.
  Ross.
\newblock Spin differences in the $^{90}\mathrm{Zr}$ compound nucleus induced by $(p,p')$, $(p,d)$, and $(p,t)$ surrogate reactions.
\newblock {\em Phys. Rev. C}, 92:054603, Nov 2015.

\bibitem{SPP-2020}
R.~P\'erez~S\'anchez, B.~Jurado, V.~M\'eot, O.~Roig, M.~Dupuis, O.~Bouland,
  D.~Denis-Petit, P.~Marini, L.~Mathieu, I.~Tsekhanovich, M.~A\"{\i}che,
  L.~Audouin, C.~Cannes, S.~Czajkowski, S.~Delpech, A.~G\"orgen, M.~Guttormsen,
  A.~Henriques, G.~Kessedjian, K.~Nishio, D.~Ramos, S.~Siem, and F.~Zeiser.
\newblock Simultaneous determination of neutron-induced fission and radiative
  capture cross sections from decay probabilities obtained with a surrogate
  reaction.
\newblock {\em Phys. Rev. Lett.}, 125:122502, Sep 2020.

\bibitem{SPP-2016}
Q.~Ducasse, B.~Jurado, M.~A\"{\i}che, P.~Marini, L.~Mathieu, A.~G\"orgen,
  M.~Guttormsen, A.~C. Larsen, T.~Tornyi, J.~N. Wilson, G.~Barreau, G.~Boutoux,
  S.~Czajkowski, F.~Giacoppo, F.~Gunsing, T.~W. Hagen, M.~Lebois, J.~Lei,
  V.~M\'eot, B.~Morillon, A.~M. Moro, T.~Renstr\o{}m, O.~Roig, S.~J. Rose,
  O.~S\'erot, S.~Siem, I.~Tsekhanovich, G.~M. Tveten, and M.~Wiedeking.
\newblock Investigation of the $^{238}\mathrm{U}(d,p)$ surrogate reaction via
  the simultaneous measurement of $\gamma$-decay and fission probabilities.
\newblock {\em Phys. Rev. C}, 94:024614, Aug 2016.

\bibitem{escher-2012-review}
Jutta~E. Escher, Jason~T. Harke, Frank~S. Dietrich, Nicholas~D. Scielzo, Ian~J.
  Thompson, and Walid Younes.
\newblock Compound-nuclear reaction cross sections from surrogate measurements.
\newblock {\em Rev. Mod. Phys.}, 84:353--397, Mar 2012.

\bibitem{Boutoux-2012}
G.~Boutoux, B.~Jurado, V.~Méot, O.~Roig, L.~Mathieu, M.~Aïche, G.~Barreau,
  N.~Capellan, I.~Companis, S.~Czajkowski, K.-H. Schmidt, J.T. Burke, A.~Bail,
  J.M. Daugas, T.~Faul, P.~Morel, N.~Pillet, C.~Théroine, X.~Derkx, O.~Sérot,
  I.~Matéa, and L.~Tassan-Got.
\newblock Study of the surrogate-reaction method applied to neutron-induced
  capture cross sections.
\newblock {\em Physics Letters B}, 712(4):319--325, 2012.

\bibitem{escher-2010-1}
N.~D. Scielzo, J.~E. Escher, J.~M. Allmond, M.~S. Basunia, C.~W. Beausang,
  L.~A. Bernstein, D.~L. Bleuel, J.~T. Harke, R.~M. Clark, F.~S. Dietrich,
  P.~Fallon, J.~Gibelin, B.~L. Goldblum, S.~R. Lesher, M.~A. McMahan, E.~B.
  Norman, L.~Phair, E.~Rodriquez-Vieitez, S.~A. Sheets, I.~J. Thompson, and
  M.~Wiedeking.
\newblock Measurement of $\gamma$-emission branching ratios for
  $^{154,156,158}\mathrm{Gd}$ compound nuclei: Tests of surrogate nuclear
  reaction approximations for $(n,\gamma$) cross sections.
\newblock {\em Phys. Rev. C}, 81:034608, Mar 2010.

\bibitem{escher-2010}
Jutta~E. Escher and Frank~S. Dietrich.
\newblock Cross sections for neutron capture from surrogate measurements: An examination of Weisskopf-Ewing and ratio approximations.
\newblock {\em Phys. Rev. C}, 81:024612, Feb 2010.

\bibitem{Chiba-2010}
Satoshi Chiba and Osamu Iwamoto.
\newblock Verification of the surrogate ratio method.
\newblock {\em Phys. Rev. C}, 81:044604, Apr 2010.

\bibitem{Forssen-2007}
C.~Forss\'en, F.~S. Dietrich, J.~Escher, R.~D. Hoffman, and K.~Kelley.
\newblock Determining neutron capture cross sections via the surrogate reaction technique.
\newblock {\em Phys. Rev. C}, 75:055807, May 2007.

\bibitem{Smith-Wambach1988}
R.~D. Smith and J.~Wambach.
\newblock Damping of the continuum response from $2p-2h$ excitations.
\newblock {\em Phys. Rev. C}, 38:100--108, Jul 1988.

\bibitem{damping-2}
M.G.E. Brand, G.A. Rijsdijk, F.A. Muller, K.~Allaart, and W.H. Dickhoff.
\newblock Fragmentation of single-particle strength and the validity of the
  shell model.
\newblock {\em Nuclear Physics A}, 531(2):253--284, 1991.

\bibitem{damping-1}
G.~F. Bertsch, P.~F. Bortignon, and R.~A. Broglia.
\newblock Damping of nuclear excitations.
\newblock {\em Rev. Mod. Phys.}, 55:287--314, Jan 1983.

\bibitem{Chimanski-2025}
E.~V. Chimanski, E.~J. In, S.~P\'eru, A.~Thapa, W.~Younes, and J.~E. Escher.
\newblock Coupling between collective modes in the deformed $^{98}\mathrm{Zr}$ nucleus: Insights from consistent $\mathrm{HFB}+\mathrm{QRPA}$ calculations
  with the gogny interaction.
\newblock {\em Phys. Rev. C}, 111:054314, May 2025.

\bibitem{Chimanski-2022}
Emanuel~V. Chimanski, Eun~Jin In, Jutta~E. Escher, Sophie Péru, and Walid
  Younes.
\newblock Towards a predictive HFB$+$QRPA framework for deformed nuclei: Selected
  tools and techniques.
\newblock {\em Journal of Physics: Conference Series}, 2340(1):012033, sep
  2022.

\bibitem{Peru2014}
Sophie P\'{e}ru and Marco Martini.
\newblock Mean field based calculations with the Gogny force: Some theoretical tools to explore the nuclear structure.
\newblock {\em The European Physical Journal A}, 50, 05 2014.

\bibitem{Peru-2011}
S.~P\'eru, G.~Gosselin, M.~Martini, M.~Dupuis, S.~Hilaire, and J.-C. Devaux.
\newblock Giant resonances in $^{238}\mathrm{U}$ within the quasiparticle
  random-phase approximation with the Gogny force.
\newblock {\em Phys. Rev. C}, 83:014314, Jan 2011.

\bibitem{jlm-1977}
J.-P. Jeukenne, A.~Lejeune, and C.~Mahaux.
\newblock Optical-model potential in finite nuclei from Reid's hard core
  interaction.
\newblock {\em Phys. Rev. C}, 16:80--96, Jul 1977.

\bibitem{jlmb-1998}
E.~Bauge, J.~P. Delaroche, and M.~Girod.
\newblock Semi-microscopic nucleon-nucleus spherical optical model for nuclei with $A\geq40$ at energies up to 200 MeV.
\newblock {\em Phys. Rev. C}, 58:1118--1145, Aug 1998.

\bibitem{Thompson1988}
Ian~J. Thompson.
\newblock Coupled reaction channels calculations in nuclear physics.
\newblock {\em Computer Physics Reports}, 7(4):167--212, 1988.

\bibitem{dupuis2019}
M.~Dupuis, G.~Haouat, J.-P. Delaroche, E.~Bauge, and J.~Lachkar.
\newblock Challenging microscopic structure and reaction models for nucleon
  scattering off nuclei in the $A=208$ mass region.
\newblock {\em Phys. Rev. C}, 100:044607, Oct 2019.

\bibitem{dupuis2017}
{Dupuis, M.}
\newblock Microscopic description of elastic and direct inelastic nucleon
  scattering off spherical nuclei.
\newblock {\em Eur. Phys. J. A}, 53(5):111, 2017.

\bibitem{2018-inelastic-jlm}
M.~L. Cort\'es, P.~Doornenbal, M.~Dupuis, S.~M. Lenzi, F.~Nowacki,
  A.~Obertelli, S.~P\'eru, N.~Pietralla, V.~Werner, K.~Wimmer, G.~Authelet,
  H.~Baba, D.~Calvet, F.~Ch\^ateau, A.~Corsi, A.~Delbart, J-M. Gheller,
  A.~Gillibert, T.~Isobe, V.~Lapoux, C.~Louchart, M.~Matsushita, S.~Momiyama,
  T.~Motobayashi, M.~Niikura, H.~Otsu, C.~P\'eron, A.~Peyaud, E.~C. Pollacco,
  J-Y. Rouss\'e, H.~Sakurai, C.~Santamaria, M.~Sasano, Y.~Shiga, S.~Takeuchi,
  R.~Taniuchi, T.~Uesaka, H.~Wang, K.~Yoneda, F.~Browne, L.~X. Chung, Zs.
  Dombradi, S.~Franchoo, F.~Giacoppo, A.~Gottardo, K.~Hadynska-Klek,
  Z.~Korkulu, S.~Koyama, Y.~Kubota, J.~Lee, M.~Lettmann, R.~Lozeva, K.~Matsui,
  T.~Miyazaki, S.~Nishimura, L.~Olivier, S.~Ota, Z.~Patel, E.~Sahin, C.~M.
  Shand, P-A. S\"oderstr\"om, I.~Stefan, D.~Steppenbeck, T.~Sumikama,
  D.~Suzuki, Zs. Vajta, J.~Wu, and Z.~Xu.
\newblock Inelastic scattering of neutron-rich Ni and Zn isotopes off a proton
  target.
\newblock {\em Phys. Rev. C}, 97:044315, Apr 2018.

\bibitem{thapa2023}
{Thapa, Aaina}, {Escher, Jutta}, {Chimanski, Emanuel}, {Dupuis, Marc}, {Péru,
  Sophie}, and {Younes, Walid}.
\newblock Predicting nucleon-nucleus scattering observables using nuclear
  structure theory.
\newblock {\em EPJ Web Conf.}, 292:06003, 2024.

\bibitem{hole_states_data1}
G.~Duhamel-Chr\'etien, G.~Perrin, C.~Perrin, V.~Comparat, E.~Gerlic,
  S.~Gal\`es, and C.~P. Massolo.
\newblock Neutron hole states in $^{89}\mathrm{Zr}$ via the $(p,d)$ reaction at 58 MeV.
\newblock {\em Phys. Rev. C}, 43:1116--1126, Mar 1991.

\bibitem{hole_states_data2}
S.~Fortier, S.~Gales, Sam~M. Austin, W.~Benenson, G.~M. Crawley, C.~Djalali,
  J.~S. Winfield, and G.~Yoo.
\newblock One-nucleon-transfer reactions induced by $^{20}\mathrm{Ne}$ at 500 and 600 MeV.
\newblock {\em Phys. Rev. C}, 41:2689--2697, Jun 1990.

\bibitem{bespalova-2015}
O.V. Bespalova, E.A. Romanovsky, and T.I. Spasskaya.
\newblock Dispersive optical potential for nuclei with $N$ and $Z$ values
  changing toward the nucleon drip lines.
\newblock {\em Phys. Atom. Nuclei}, 78:118–127, February 2015.

\bibitem{netterdon-2015}
L.~Netterdon, A.~Endres, S.~Goriely, J.~Mayer, P.~Scholz, M.~Spieker, and
  A.~Zilges.
\newblock Experimental constraints on the $\gamma$-ray strength function in $^{90}$Zr using partial cross sections of the $^{89}$Y$(p,\gamma)^{90}$Zr reaction.
\newblock {\em Physics Letters B}, 744:358--362, 2015.

\bibitem{Fedorets_2013}
I.~D. Fedorets and S.~S. Ratkevich.
\newblock Radiative strength functions for dipole transitions in $^{90}$Zr.
\newblock {\em Physics of Atomic Nuclei}, 76(1):44–55, January 2013.

\bibitem{PDR-2008}
R.~Schwengner, G.~Rusev, N.~Tsoneva, N.~Benouaret, R.~Beyer, M.~Erhard,
  E.~Grosse, A.~R. Junghans, J.~Klug, K.~Kosev, H.~Lenske, C.~Nair, K.~D.
  Schilling, and A.~Wagner.
\newblock Pygmy dipole strength in $^{90}\mathrm{Zr}$.
\newblock {\em Phys. Rev. C}, 78:064314, Dec 2008.

\bibitem{GS-1979}
G.~Szeflinska, Z.~Szefliński, and Z.~Wilhelmi.
\newblock Gamma-ray strength functions for $A = $ 70–90 nuclei.
\newblock {\em Nuclear Physics A}, 323(2):253--270, 1979.

\bibitem{Y89pg-2}
S.~Harissopulos, A.~Spyrou, A.~Lagoyannis, M.~Axiotis, P.~Demetriou, J.~W.
  Hammer, R.~Kunz, and H.-W. Becker.
\newblock Cross section measurements of proton capture reactions relevant to
  the $p$ process: The case of ${}^{89}$Y${(p,\ensuremath{\gamma})}^{90}$Zr and
  ${}^{121,123}$Sb($p,\ensuremath{\gamma}$)${}^{122,124}$Te.
\newblock {\em Phys. Rev. C}, 87:025806, Feb 2013.

\bibitem{Y89pg-3}
P.~Tsagkari, E.~Skreti, G.~A. Souliotis, P.~Demetriou, S.~Harissopulos,
  T.~Paradellis, J.~W. Hammer, R.~Kunz, C.~Angulo, S.~Goriely, and T.~Rauscher.
\newblock Cross section measurements of the $^{89}$Y$(p,\gamma)^{90}$Zr reaction in the energy range $E_p$=1.6-2.4 MeV.
\newblock {\em HNPS Advances in Nuclear Physics}, 10:154–159, Dec. 2019.

\bibitem{YAHFC}
William~E. Ormand.
\newblock Yet another Hauser-Feshbach code.
\newblock {Computer Software, https://doi.org/10.11578/dc.20211209.1}, Jul 2021.

\bibitem{PT-original}
C.~E. Porter and R.~G. Thomas.
\newblock Fluctuations of nuclear reaction widths.
\newblock {\em Phys. Rev.}, 104:483--491, Oct 1956.

\bibitem{gorton-2025}
{Gorton, Oliver} and {Escher, Jutta}.
\newblock Correcting for neutron width fluctuations in Hauser-Feshbach gamma branching ratios.
\newblock {\em EPJ Web Conf.}, 322:02004, 2025.

\bibitem{PT-dicebox}
M.~Krti\ifmmode~\check{c}\else \v{c}\fi{}ka, S.~Goriely, S.~Hilaire, S.~P\'eru,
  and S.~Valenta.
\newblock Constraints on the dipole photon strength functions from experimental multistep cascade spectra.
\newblock {\em Phys. Rev. C}, 99:044308, Apr 2019.

\bibitem{PT-exact}
J.~L. Tain, E.~Valencia, A.~Algora, J.~Agramunt, B.~Rubio, S.~Rice,
  W.~Gelletly, P.~Regan, A.-A. Zakari-Issoufou, M.~Fallot, A.~Porta,
  J.~Rissanen, T.~Eronen, J.~\"Ayst\"o, L.~Batist, M.~Bowry, V.~M. Bui,
  R.~Caballero-Folch, D.~Cano-Ott, V.-V. Elomaa, E.~Estevez, G.~F. Farrelly,
  A.~R. Garcia, B.~Gomez-Hornillos, V.~Gorlychev, J.~Hakala, M.~D. Jordan,
  A.~Jokinen, V.~S. Kolhinen, F.~G. Kondev, T.~Mart\'{\i}nez, E.~Mendoza,
  I.~Moore, H.~Penttil\"a, Zs. Podoly\'ak, M.~Reponen, V.~Sonnenschein, and
  A.~A. Sonzogni.
\newblock Enhanced $\gamma$-ray emission from neutron unbound
  states populated in $\beta$ decay.
\newblock {\em Phys. Rev. Lett.}, 115:062502, Aug 2015.

\bibitem{PT-WIP}
A.~Thapa~et al.
\newblock Porter-Thomas fluctuations in surrogate reaction applications, (unpublished results).

\bibitem{Kopecky-1990}
J.~Kopecky and M.~Uhl.
\newblock Test of gamma-ray strength functions in nuclear reaction model
  calculations.
\newblock {\em Phys. Rev. C}, 41:1941--1955, May 1990.

\bibitem{gilbert-1965}
A.~Gilbert and A.~G.~W. Cameron.
\newblock A composite nuclear-level density formula with shell corrections.
\newblock {\em Canadian Journal of Physics}, 43:1446--1496, 1965.

\bibitem{commcas-2025}
O.C. Gorton and J.E. Escher.
\newblock Commcas: Computational model monte carlo sampler.
\newblock Technical Report LLNL-TR-2011377, Lawrence Livermore National
  Laboratory, 2025.

\bibitem{gorton2020neutron}
Oliver Gorton and Jutta~E. Escher.
\newblock {Neutron Capture Cross Sections from Surrogate Reaction Data and Theory: Connecting the Pieces with a Markov-Chain Monte Carlo Approach}, page 229–231.
\newblock Springer International Publishing, September 2020.

\bibitem{goodman2010ensemble}
Jonathan Goodman and Jonathan Weare.
\newblock Ensemble samplers with affine invariance.
\newblock {\em Communications in Applied Mathematics and Computational
  Science}, 5(1):65–80, January 2010.

\bibitem{PDR-2012}
C.~Iwamoto, H.~Utsunomiya, A.~Tamii, H.~Akimune, H.~Nakada, T.~Shima,
  T.~Yamagata, T.~Kawabata, Y.~Fujita, H.~Matsubara, Y.~Shimbara, M.~Nagashima,
  T.~Suzuki, H.~Fujita, M.~Sakuda, T.~Mori, T.~Izumi, A.~Okamoto, T.~Kondo,
  B.~Bilgier, H.~C. Kozer, Y.-W. Lui, and K.~Hatanaka.
\newblock Separation of pygmy dipole and $M1$ resonances in $^{90}\mathrm{Zr}$ by a high-resolution inelastic proton scattering near 0\ifmmode^\circ\else\textdegree\fi{}.
\newblock {\em Phys. Rev. Lett.}, 108:262501, Jun 2012.

\bibitem{M1-2013}
G.~Rusev, N.~Tsoneva, F.~D\"onau, S.~Frauendorf, R.~Schwengner, A.~P. Tonchev,
  A.~S. Adekola, S.~L. Hammond, J.~H. Kelley, E.~Kwan, H.~Lenske, W.~Tornow,
  and A.~Wagner.
\newblock Fine structure of the giant $M1$ resonance in $^{90}\mathrm{Zr}$.
\newblock {\em Phys. Rev. Lett.}, 110:022503, Jan 2013.

\bibitem{M1-data-1980}
D.~Meuer, R.~Frey, D.H.H. Hoffmann, A.~Richter, E.~Spamer, O.~Titze, and
  W.~Knüpfer.
\newblock High resolution inelastic electron scattering on $^{90}$Zr at low momentum transfer and strong fragmentation of the magnetic quadrupole strength.
\newblock {\em Nuclear Physics A}, 349(3):309--338, 1980.

\bibitem{OMP-pg-senstivity}
E.~Vagena, M.~Axiotis, and P.~Dimitriou.
\newblock Systematics of the semi-microscopic proton-nucleus optical potential at low energies relevant for nuclear astrophysics.
\newblock {\em Phys. Rev. C}, 103:045806, Apr 2021.

\bibitem{JENDL5}
Osamu Iwamoto, Nobuyuki Iwamoto, Satoshi Kunieda, Futoshi Minato, Shinsuke
  Nakayama, Yutaka Abe, Kohsuke Tsubakihara, Shin Okumura, Chikako Ishizuka,
  Tadashi Yoshida, Satoshi Chiba, Naohiko Otuka, Jean-Christophe Sublet, Hiroki
  Iwamoto, Kazuyoshi Yamamoto, Yasunobu Nagaya, Kenichi Tada, Chikara Konno,
  Norihiro Matsuda, Kenji Yokoyama, Hiroshi Taninaka, Akito Oizumi, Masahiro
  Fukushima, Shoichiro Okita, Go~Chiba, Satoshi Sato, Masayuki Ohta, and Saerom
  Kwon.
\newblock Japanese evaluated nuclear data library version 5: JENDL-5.
\newblock {\em Journal of Nuclear Science and Technology}, 60(1):1--60, 2023.

\bibitem{JEFF3}
Plompen A, Cabellos O, De~Saint~Jean C, Fleming M, Algora A, Angelone M,
  Archier P, Bauge E, Bersillon O, Blokhin A, Cantargi F, Chebboubi A, Diez CJ,
  Duarte H, Dupont E, Dyrda J, Erasmus B, Fiorito L, Fischer U, Flammini D,
  Foligno D, Gilbert M, Granada JR, Haeck W, Hambsch F, Helgesson P, Hilaire S,
  Hill I, Hursin M, Ichou R, Jacqmin R, Jansky B, Jouanne C, Kellett M, Kim DH,
  Kim HI, Kodeli I, Koning AJ, Konobeyev AY, Kopecky S, Kos B, Krasa A, Leal L,
  Leclaire N, Leconte P, Lee YO, Leeb H, Litaize O, Majerle M, Marquez~Damian
  J, Michel-Sendis F, Mills R, Morillon B, Noguere G, Pecchia M, Pelloni S,
  Pereslavtsev P, Perry R, Rochman D, Roehrmoser R, Romain P, Romojaro P,
  Roubtsov D, Sauvan P, Schillebeeckx P, Schmidt K, Serot O, Simakov S, Sirakov
  I, Sjöstrand H, Stankovskiy A, Sublet JC, Tamagno P, Trkov A, Van Den~Marck
  S, Velarde F, Villari R, Yokoyama K, and Zerovnik G.
\newblock The joint evaluated fission and fusion nuclear data library,
  JEFF-3.3.
\newblock {\em European Physical Journal A}, 56:181, 2020.

\bibitem{TENDL-2023}
D.~Rochman, A.~Koning, S.~Goriely, and S.~Hilaire.
\newblock Tendl-astro: A new nuclear data set for astrophysics interest.
\newblock {\em Nuclear Physics A}, 1053:122951, 2025.

\bibitem{goriely2025}
S.~Goriely, S.~Péru, and S.~Hilaire.
\newblock QRPA prediction of the nuclear level densities and de-excitation
  photon strength functions.
\newblock {\em Physics Letters B}, 868:139677, 2025.

\bibitem{M1-shell-model-2025}
Fang-Qi Chen, Y.~F. Niu, Yang Sun, and Mathis Wiedeking.
\newblock Origin of the low-energy enhancement of the $\gamma$-ray
  strength function.
\newblock {\em Phys. Rev. Lett.}, 134:082502, Feb 2025.

\bibitem{M1-shell-model-2022}
S.~Frauendorf and R.~Schwengner.
\newblock Evolution of low-lying {{M1}} modes in germanium isotopes.
\newblock {\em Physical Review C: Nuclear Physics}, 105(3):034335, March 2022.

\bibitem{M1-shell-model-2018}
J.~E. Midtb{\o}, A.~C. Larsen, T.~Renstr{\o}m, F.~L. Bello~Garrote, and
  E.~Lima.
\newblock Consolidating the concept of low-energy magnetic dipole decay
  radiation.
\newblock {\em Physical Review C: Nuclear Physics}, 98(6):064321, December
  2018.

\bibitem{M1-shell-model-2017}
K.~Sieja.
\newblock Electric and magnetic dipole strength at low energy.
\newblock {\em Phys. Rev. Lett.}, 119:052502, Jul 2017.

\bibitem{M1-shell-model-2013}
R.~Schwengner, S.~Frauendorf, and A.~C. Larsen.
\newblock Low-energy enhancement of magnetic dipole radiation.
\newblock {\em Physical Review Letters}, 111(23):232504, December 2013.

\end{thebibliography}






\end{document}